\documentclass[twocolumn,showpacs,preprintnumbers,amsmath,amssymb]{revtex4}

\usepackage{graphicx}
\usepackage{textcomp}

\begin{document}
\draft
\title{Optical realization of relativistic non-Hermitian quantum mechanics}
  \normalsize

\author{Stefano Longhi}
\address{Dipartimento di Fisica, Politecnico di Milano, Piazza L. da Vinci 32, I-20133 Milano, Italy}


%
\bigskip
\begin{abstract}
\noindent Light propagation in distributed feedback optical
structures with gain/loss regions is shown to provide an accessible
laboratory tool to visualize in optics the spectral properties of
the one-dimensional Dirac equation with non-Hermitian interactions.
Spectral singularities and $\mathcal{P}\mathcal{T}$ symmetry
breaking of the Dirac Hamiltonian are shown to correspond to simple
observable physical quantities  and related to well-known physical
phenomena like resonance narrowing and laser oscillation.
\end{abstract}


\maketitle

{\it Introduction.} Quantum mechanics prescribes that the
Hamiltonian $H$ of a physical system must be self-adjoint. Since the
the seminal paper by Bender and Boettcher \cite{Bender98}, it was
realized that the Hermiticity of $H$ can be relaxed, and that a a
consistent quantum theory can be constructed for a broader class of
Hamiltonians
\cite{Bender02,Mostafazadeh02,BenderReview,MostafazadehReview}.
Among these are parity-time ($\mathcal{PT}$) Hamiltonians, which
possess a real-valued spectrum below a symmetry-breaking point.
Non-Hermitian Hamiltonians are also
 encountered in reduced descriptions of open quantum systems, with important applications
to atomic, molecular and condensed-matter physics
\cite{RotterReview}. In such systems, the lack of Hermiticity can
lead to the appearance of exceptional points and spectral
singularities, whose physical relevance has been discussed by
several authors (see, e.g., \cite{Berry,O3,O4}). Recently,
non-Hermitian extensions of relativistic wave equations
\cite{D1,D2,D3} and non-Hermitian quantum field theories
\cite{field} have attracted an increasing interest as well. As some
issues in this field are still debated (see, e.g., \cite{field}),
physical realizations of non-Hermitian relativistic models remain
mostly unexplored. Recently, optical structures in media with a
complex refractive index have been proposed to test and visualize
non-Hermitian features rooted in
 the non-relativistic Schr\"{o}dinger
 equation with a complex potential \cite{O2,O3,O1,O4}. The main motivation in the study of such quantum-optical analogs
 is that concepts like exceptional points, spectral
singularities and $\mathcal{PT}$ symmetry breaking become measurable
quantities in an optical experiment \cite{O3,O4}. This has lead to
the first experimental visualization of exceptional points and
$\mathcal{PT}$ symmetry breaking in an optical structure
\cite{Salamo09,Ruter10}. Such results motivate the search for
optical simulators of non-Hermitian {\it relativistic} wave
equations, which is the aim of this Letter. Here it is shown that
light propagation in distributed-feedback (DFB) optical structures
with gain and/or loss regions, which is at the heart of such
important devices as DFB semiconductor lasers \cite{Shank,Poladian},
can provide a fertile ground to test the spectral properties of
non-Hermitian Dirac Hamiltonians. Similarities between light
propagation in DFB structures and relativistic wave equations were
noticed in early studies on gap solitons in connection with the
massive Thirring model of field theory \cite{note0}, however these
previous studies did not consider non-Hermitian
interactions.\\
{\it Non-Hermitian Dirac equation and its optical realization.} Let
us consider the Dirac equation in one spatial dimension for a
two-component spinor wave function $\psi(x,t)=(\psi_1,\psi_2)^T$
with time-independent vector ($V$) and scalar ($S$) couplings, which
in natural units ($\hbar=c=1$) reads \cite{D2,D3}
\begin{equation}
i \partial_t \psi= -i \alpha \partial_x \psi+ \beta m(x) \psi+V(x)
\psi \equiv H \psi
\end{equation}
where $\alpha$ and $\beta$ are $2 \times 2$ Hermitian square
matrices satisfying the relations $\alpha^2=\beta^2=1$ and $\alpha
\beta+\beta \alpha=0$, $V(x)$ is the time-component of a Lorentzian
2-vector potential, $m(x)$ is the space-dependent effective mass
defined by $m(x)=m_0+S(x)$, and $m_0$ is the rest mass of the Dirac
particle. Among the various representations of the Dirac equation,
the optical realization of Eq.(1) discussed below is at best
highlighted by taking $\alpha=\sigma_z$ and
$\beta=\sigma_x$, where $\sigma_x= \left( \begin{array}{cc} 0 & 1 \\
1 & 0 \end{array} \right)$ and $\sigma_z= \left( \begin{array}{cc} 1& 0 \\
0 & -1 \end{array} \right)$ are the Pauli matrices. If either $V$ or
$m$ assume complex values, the Dirac Hamiltonian $H$ is
non-Hermitian and its spectrum is, in general, complex-valued. For
the Dirac equation, parity and time-reversal operators can be
defined as \cite{D3} $\mathcal{P}= \mathcal{P}_0 \beta$ and
$\mathcal{T}= \mathcal{K} \beta$, where $\mathcal{P}_0$ changes $x$
with $-x$ and $\mathcal{K}$ performs complex-conjugation. Hence
$\mathcal{PT}\psi(x)=\psi^*(-x)$.  $\mathcal{PT}$ invariance of $H$,
i.e. $[\mathcal{PT},H]=0$, requires $V(-x)=V^*(x)$ and
$m(-x)=m^*(x)$ \cite{D3}. If every eigenfunction of a
$\mathcal{PT}$-invariant Hamiltonian is also an eigenfunction of the
$\mathcal{PT}$ operator, the $\mathcal{PT}$ symmetry of $H$ is said
to be unbroken and its spectrum is real-valued \cite{BenderReview}.
In the following, we will consider a real-valued effective mass
$m(x)$, whereas non-Hermiticity is introduced by allowing the vector
potential $V(x)$ to be complex-valued. An optical realization of the
Dirac equation (1) is provided by propagation of light waves in an
effective one-dimensional DFB structure \cite{Shank}. Let
$n(z)=n_0-\Delta n h(z) \cos(2 \pi z/ \Lambda+2 \theta(z))$ be the
effective index grating of the dielectric structure, where $n_0$ is
the modal refractive index in absence of the grating, $\Delta n \ll
n_0$ and $\Lambda$ are the peak index change and the nominal period
of the grating, respectively, and $h(z)$, $2 \theta(z)$ are the
normalized amplitude and phase profiles, respectively, of the index
grating. The periodic modulation of the refractive index leads to
Bragg scattering between two counterpropagating waves at frequencies
close to the Bragg frequency $\omega_B=\pi c/(\Lambda n_0)$, where
$c$ is the speed of light in vacuum. The linear space-dependent
absorption coefficient of counterpropagating waves in the structure
is indicated by $ \alpha_0 (x)$ ($\alpha_0 >0$ in lossy regions,
$\alpha_0<0$ in gain regions). In a semiconductor DFB structure,
gain and loss regions could be tailored by a judicious control of
current injection across the active layer \cite{Poladian}.
Indicating by $E(z,\tau)=\psi_1(z,\tau) \exp[-i \omega_B \tau +ik_B
z+i \theta(z)]+ \psi_2(z,\tau) \exp[-i \omega_B \tau -ik_B z-i
\theta(z)]+c.c.$ the electric field propagating in the DFB
structure, where $k_B=\pi/\Lambda$, the envelopes $\psi_1$ and
$\psi_2$ of counterpropagating waves satisfy coupled-mode equations
\cite{Poladian}. After introduction of the scaled space and time
variables $x=z/Z$ and $t=\tau/T$, with $Z=2n_0 \Lambda/(\pi \Delta
n)$ and $T=Z/v_g$, where $v_g \simeq c/n_0$ is the group velocity of
light at frequency $\omega_B$, the envelopes $\psi_1$ and $\psi_2$
satisfy Eq.(1) with a real mass $m$ and a complex-valued vector
potential $V$ defined by
\begin{equation}
m(x)=h(x) , \; \;  V(x)=\frac{d \theta}{dx}-i \gamma(x),
\end{equation}
where $\gamma(x)=Z \alpha_0 (x)$ is the dimensionless absorption
coefficient. In the following, we will assume that both $m(x)$ and
$V(x)$ have a limited support, i.e. $m=V=0$ for $|x|>L/2$. This case
typically applies to optical DFB structures, in which the grating
region is spatially confined to a finite region
of length $L$.\\
{\it Spectral singularities, bound states and resonances of the
Dirac Hamiltonian.} The spectral properties of the non-Hermitian
Dirac Hamiltonian $H$, defined by Eqs.(1) and (2), can be
investigated by standard methods of scattering theory. To this aim,
let us introduce the functions $\phi^{(1)}_E(x)$, $\phi^{(2)}_E(x)$,
$\varphi^{(1)}_E(x)$ and $\varphi^{(2)}_E(x)$, which satisfy the
equation $H \psi=E \psi$ ($E$ is a complex-valued parameter) with
the asymptotic behavior $\phi^{(1)}_E=(1,0)^T \exp(iEx)$,
$\phi^{(2)}_E=(0,1)^T \exp(-iEx)$ for $x<-L/2$, and
$\varphi^{(1)}_E=(1,0)^T \exp(iEx)$, $\varphi^{(2)}_E=(0, 1)^T
\exp(-iEx)$ for $x>L/2$. As the Wronskians $W\{
\phi^{(1)}_E,\phi^{(2)}_E \}=W \{ \varphi^{(1)}_E,\varphi^{(2)}_E
\}=1$ do not vanish, $\{ \phi^{(1)}_E, \phi^{(2)}_E \}$ and $\{
\varphi^{(1)}_E, \varphi^{(2)}_E \}$ are two sets of
linearly-independent solutions to the equation $(E-H)\psi=0$, and
therefore there exists a $ 2 \times 2$ matrix $\mathcal{M}(E)$, with
$\mathrm{det} \mathcal{M}(E)=1$, such that
\begin{eqnarray}
\phi_{E}^{(1)} (x) & = & \mathcal{M}_{11}(E)
\varphi_{E}^{(1)}(x)+\mathcal{M}_{21}(E)
\varphi_{E}^{(2)}(x) \nonumber \\
\phi_{E}^{(2)} (x) & = & \mathcal{M}_{12}(E)
\varphi_{E}^{(1)}(x)+\mathcal{M}_{22}(E) \varphi_{E}^{(2)}(x).
\end{eqnarray}
Physically, $\mathcal{M}(E)$ is the transfer matrix that connects
the amplitudes of forward- and backward-propagating waves from
$x=-L/2$ to $x=L/2$, which is commonly adopted in the optical
context (see, for instance, \cite{Poladian}). The spectral
transmission and reflection coefficients, for left ($l$) and right
($r$) incidence, can be expressed in terms of the transfer matrix
elements in the usual way \cite{O4}
\begin{equation}
t^{(l)}=t^{(r)}\equiv t =\frac{1}{\mathcal{M}_{22}} \; , \; \;
r^{(l)}=-\frac{\mathcal{M}_{21}}{\mathcal{M}_{22}} \; , \; \;
r^{(r)}=\frac{\mathcal{M}_{12}}{\mathcal{M}_{22}}.
\end{equation}
For a $\mathcal{PT}$-invariant Hamiltonian, one has
$\varphi_{E^*}^{(1)}(x)=\phi_{E}^{(1)*}(-x)$ and
$\varphi_{E^*}^{(2)}(x)=\phi_{E}^{(2)*}(-x)$, and thus
$\mathcal{M}^{-1}(E)=\mathcal{M}^{*}(E^*)$, which implies
$\mathcal{M}_{11}(E)=\mathcal{M}_{22}^*(E^*)$.\\
The spectrum of $H$, as well as the existence of spectral
singularities arising from the non-Hermiticity of $H$
\cite{O4,Naimark}, can be determined by an inspection of the
singularities and branch cuts of the resolvent $G(E)=(E-H)^{-1}$,
which takes the integral form \cite{Naimark} $G(E) \psi(x)= \int dy
\mathcal{G}(x,y;E)\psi(y)$, where $\mathcal{G}(x,y;E)$ is the Green
function. Its explicit form reads
$\mathcal{G}(x,y;E)=\mathcal{G}_+(x,y;E)$ for
 $\mathrm{Im}(E)>0$ and $\mathcal{G}(x,y;E)=\mathcal{G}_-(x,y;E)$ for
 $\mathrm{Im}(E)<0$, where
\begin{eqnarray}
\mathcal{G}_+(x,y;E)=-\frac{i}{\mathcal{M}_{22}(E)} \left[ \Phi(y-x)
\phi^{(2)}_E (x)\varphi^{(1)T}_E (y)+ \right. \nonumber \\ \left.
\Phi(x-y) \varphi^{(1)}_E (x) \phi^{(2)T}_E (y) \right] \sigma_x,
\end{eqnarray}
\begin{eqnarray}
\mathcal{G}_-(x,y;E)=\frac{i}{\mathcal{M}_{11}(E)} \left[ \Phi(y-x)
\phi^{(1)}_E (x)\varphi^{(2)T}_E (y)+ \right. \nonumber \\ \left.
\Phi(x-y) \varphi^{(2)}_E (x) \phi^{(1)T}_E (y) \right] \sigma_x,
\end{eqnarray}
and $\Phi(x)$ is the Heaviside function [$\Phi(x)=0$ for $x<0$,
$\Phi(x)=1$ for $x>0$]. On the basis of Eqs.(5) and (6), the
following results hold for the spectral properties of $H$.\\
(i) {\it Continuous spectrum}. The continuous spectrum of $H$ is the
entire real energy axis ($-\infty<E<\infty$), where $G(E)$ has a branch cut.\\
(ii) {\it Point spectrum}. The zeros of $\mathcal{M}_{22}(E)$ on the
$\mathrm{Im}(E)>0$ half plane, together with the zeros of
$\mathcal{M}_{11}(E)$ on the $\mathrm{Im}(E)<0$ half plane, define
the point spectrum of $H$; at such energies, the function
$\phi^{(2)}_{E}(x)$ [for $\mathrm{Im}(E)>0$] and $\phi^{(1)}_{E}(x)$
[for $\mathrm{Im}(E)<0$] are bound states of $H$.\\
(iii) {\it Resonances}. The zeros of $\mathcal{M}_{22}(E)$
[$\mathcal{M}_{11}(E)$] on the $\mathrm{Im}(E)<0$
[$\mathrm{Im}(E)>0$] half plane, i.e. the poles of the analytic
continuation of $\mathcal{G}_+$ [$\mathcal{G}_-$] on the
$\mathrm{Im}(E)<0$ [$\mathrm{Im}(E)>0$]
half plane, correspond to the resonances [antiresonances] of the scattering problem.\\
(iv) {\it Spectral singularities}. A spectral singularity
\cite{Naimark} is any zero $E=E_0$ on the real axis of either
$\mathcal{M}_{11}(E)$ or $\mathcal{M}_{22}(E)$, around which $G(E)$
is unbounded (for either $E=E_0+i0^+$ or $E=E_0-i0^+$) in spite of
the fact that $E=E_0$ does not belong to
the point spectrum of $H$.\\
Note that, for a $\mathcal{PT}$-invariant Hamiltonian,
$\mathcal{M}_{11}(E)=\mathcal{M}^{*}_{22}(E^*)$ and hence at a
spectral singularity {\it both} $\mathcal{M}_{11}$ and
$\mathcal{M}_{22}$ vanish, the resolvent is unbounded for $E=E_0 \pm
i0^+$, and the transmission and reflection spectral coefficients
diverge according to Eq.(4) \cite{note1}. For such a reason, in Ref.
\cite{O4} spectral singularities were identified with zero-width
resonances. However, for a $\mathcal{PT}$-non-invariant Hamiltonian,
a spectral singularity could arise from the vanishing of
$\mathcal{M}_{11}$ but not of $\mathcal{M}_{22}$, i.e. it might not
correspond to a divergence of the spectral transmission or
reflection coefficients. Such a case was not noticed in
Ref.\cite{O4} and has a different physical meaning: at such a
spectral singularity the potential becomes reflectionless for
simultaneous excitation with two waves of same energy $E_0$ and
amplitudes $1$ (for the wave incident from the left side) and
$\mathcal{M}_{21}$ (for the wave incident from the right side). An
example of such a spectral singularity, arising from the crossing of
an antiresonance with the
real energy axis, is discussed in the first example below.\\
\begin{figure}
\includegraphics[scale=0.4]{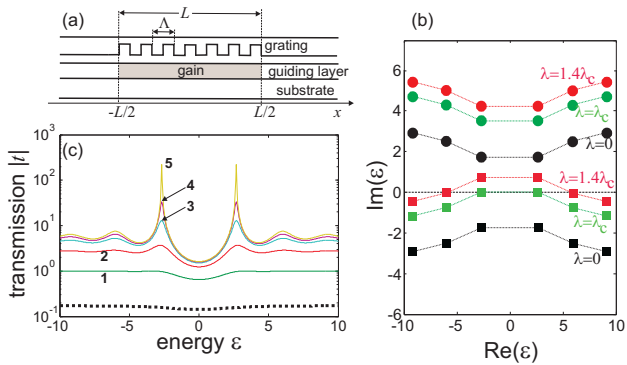}
\caption{(color online) (a) Schematic of an optical DFB structure
consisting of a uniform index grating with a homogeneous gain region
that realizes a $\mathcal{PT}$-non-invariant Dirac Hamiltonian. (b)
Zeros of $\mathcal{M}_{22}$ (squares) and of $\mathcal{M}_{11}$
(circles) in the complex energy plane $\epsilon=EL$ for $m_0L=1$ and
for increasing values of normalized gain $\lambda$. The critical
value $\lambda_c$, above which bound states emerge and the spectrum
of $H$ ceases to be real-valued, is $\lambda_c \simeq 1.755/L$. (c)
Spectral transmission $|t|$ versus normalized energy $\epsilon$ of
incident wave for increasing values of $\lambda L$: Curve 1,
$\lambda L=0$; curve 2, $\lambda L=1$; curve 3, $\lambda L=1.5$;
curve 4, $\lambda L=1.65$; curve 5, $\lambda L=1.74$. The dotted
curve in (c) is the spectral transmission of the DFB structure for
$\lambda L=1.74$ when the gain region is replaced by a lossy
region.}
\end{figure}
\begin{figure}
\includegraphics[scale=0.4]{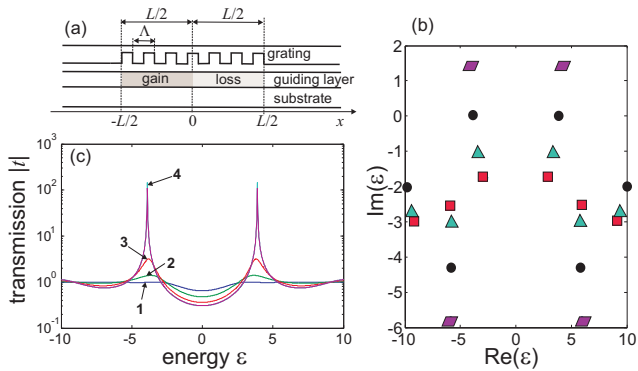}
\caption{(color online) (a) Schematic of an optical DFB structure
consisting of a uniform index grating with two homogeneous gain and
lossy regions that realizes a $\mathcal{PT}$-invariant Dirac
Hamiltonian. (b) Zeros of $\mathcal{M}_{22}$ in the complex energy
plane $\epsilon=EL$ for $m_0L=1$ and for increasing values of
$\lambda$ (squares: $\lambda=0$; triangles: $\lambda L=3$; circles:
$\lambda L=4.46$; rhombs: $\lambda L=6$). $\mathcal{PT}$ symmetry
breaking is reached at $\lambda_c \simeq 4.46/L$. (c) Behavior of
spectral transmission $|t|$  for increasing values of $\lambda L$:
Curve 1, $\lambda L=0$; curve 2, $\lambda L=3$; curve 3, $\lambda
L=4$; curve 4, $\lambda L=4.45$.}
\end{figure}
{\it Optical realizations of spectral singularities and
$\mathcal{PT}$ symmetry breaking.} Let us specialize the general
results of the spectral theory by considering two examples of
complex potentials, in which spectral singularities and
$\mathcal{PT}$ symmetry breaking of the Dirac equation correspond to
well-known physical phenomena in the theory of DFB
optical systems.\\
As a first example, let us consider a $\mathcal{PT}$-non-invariant
potential corresponding to $\theta(x)=0$ and $m(x)=m_0$,
$\gamma(x)=-\lambda$ in the interval $|x|<L/2$ [Fig.1(a)], where
$\lambda$ is the dimensionless gain coefficient. This case
corresponds to the simplest version of a DFB laser with a uniform
index grating and a uniform gain region \cite{Poladian}. The
transfer matrix of this structure is given by \cite{Poladian}
\begin{equation}
\mathcal{M}= \left(
\begin{array}{cc}
\cosh(\rho L)-i\frac{\sigma}{\rho} \sinh( \rho L ) &  -i
\frac{m_0}{\rho}
\sinh( \rho L ) \\
i\frac{m_0}{\rho} \sinh(\rho L ) & \cosh(\rho
L)+i\frac{\sigma}{\rho} \sinh( \rho L )
\end{array}
\right)
\end{equation}
where $\sigma \equiv -E+i \lambda$ and
$\rho=(m_0^2-\sigma^2)^{1/2}$. Note that the functional dependence
of $\mathcal{M}$ on $E$ and $\lambda$ is solely via
$\sigma=-E+i\lambda$, so that
$\mathcal{M}_{ik}(E,\lambda)=\mathcal{M}_{ik}(E-i\lambda,0)$. In the
Hermitian limit $\lambda=0$, the spectrum of $H$ is purely
continuous, and a number of resonances (i.e. zeros of
$\mathcal{M}_{22}$ in the $\mathrm{Im}(E)<0$ half plane) as well as
of anti-resonances (i.e. zeros of $\mathcal{M}_{11}$ in the
$\mathrm{Im}(E)>0$ half plane) do exist [Fig.1(b)]. As $\lambda$ is
increased, the spectrum remains real-valued, the resonances
(antiresonances) rigidly shift parallel to the imaginary axis, until
at a critical value $\lambda=\lambda_c$ two resonances cross the
real energy axis, i.e. they become two spectral singularities
[Fig.1(b)]. The critical value $\lambda_c$ can be expressed in the
form $\lambda_c=f(m_0L)/L$, where the function $f(m_0L)$ can be
calculated numerically. For instance, for the case of Fig.1
($m_0L=1$) one has $\lambda_c L \simeq 1.755$. At
$\lambda>\lambda_c$ the resonances cross the real energy axis and
bound states  with $\mathrm{Im}(E)>0$ emerge; correspondingly, the
spectrum of $H$ ceases to be real-valued. Such a transition, from
$\lambda<\lambda_c$ to $\lambda>\lambda_c$, is accompanied by a
narrowing of the resonance widths in the transmission spectrum as
$\lambda \rightarrow \lambda_c^-$ [see Fig.1(c)] and to the
threshold for self-oscillation at $\lambda=\lambda_c$. Note that the
imaginary parts of the eigenvalues $E$ for $\lambda>\lambda_c$ are
precisely the growth rates of the two detuned unstable modes at the
onset of lasing found in the theory of DFB lasers with a uniform
grating \cite{Shank,Poladian}. It is remarkable that the well-known
physical phenomenon of self-oscillation in DFB lasers mimics the
onset of a spectral singularity of a non-Hermitian relativistic wave
equation. Let us consider now the same DFB structure of Fig.1(a) but
with $\gamma(x)=\lambda$, corresponding to a uniform grating with a
homogeneous lossy region. The transfer matrix of the structure is
given again by Eq.(7), but with $\sigma=-E-i\lambda$. In this case,
as the loss coefficient $\lambda$ is increased from zero, the
spectrum remains real-valued shifting parallel to the imaginary
axis. At the critical value $\lambda=\lambda_c$, two antiresonances
(rather than resonances) now cross the real energy axis, i.e. they
become two spectral singularities. As opposed to the previous case,
at $\lambda=\lambda_c$ the spectral transmission does not diverge
because the spectral singularities are born from antiresonances
(rather than from resonances); see the dotted curve in Fig.1(c).
Moreover, the crossing does not correspond, as in the previous case,
to the onset of lasing in the DFB structure, because at
$\lambda>\lambda_c$ the two bound states supported by the
Hamiltonian have now a negative growth rate ($\mathrm{Im}(E)<0$),
i.e. any initial perturbation (starting from spontaneous emission
noise) is damped rather than amplified.\\
As a second example, let us propose a $\mathcal{PT}$-invariant DFB
structure which could be used to observe $\mathcal{PT}$ symmetry
breaking of the Dirac equation. The structure, schematically shown
in Fig.2(a), is composed by a uniform index grating with two
symmetric homogeneous gain and lossy regions, i.e. $\theta(x)=0$,
$m(x)=m_0$ for $|x|<L/2$, $\gamma(x)=-\lambda$ for $-L/2<x<0$, and
$\gamma(x)=\lambda$ for $0<x<L/2$. The transfer matrix of the
structure can be calculated as $\mathcal{M}=\mathcal{M}_2
\mathcal{M}_1$, where $\mathcal{M}_1$ and $\mathcal{M}_2$ are the
transfer matrices of the uniform gain and lossy sections [see
Eq.(7)]. Figure 2(b) shows the loci of zeros of $\mathcal{M}_{22}$
in the complex energy plane for increasing values of $\lambda L$ and
for $m_0L=1$. Note that, owing to the $\mathcal{PT}$-invariance of
$H$, $\mathcal{M}_{11}(E)=\mathcal{M}^{*}_{22}(E^*)$, and thus the
zeros of $\mathcal{M}_{11}$ (not shown in the figure) are simply the
complex conjugates of those of $\mathcal{M}_{22}$. As $\lambda$ is
increased above the critical value $\lambda_c \simeq 4.46/L$, a
$\mathcal{PT}$ symmetry breaking occurs, with the appearance of two
pairs of complex-conjugate eigenvalues belonging to the point
spectrum of $H$ and arising from two couples of resonances  and
antiresonances crossing the real energy axis. As the point of
$\mathcal{PT}$ symmetry breaking is approached, narrowing of the
transmission resonances is observed [see Fig.2(c)], and the onset of
lasing at $\lambda=\lambda_c^+$ corresponds to the breaking of the
$\mathcal{PT}$ symmetry. \\
{\it Conclusions}. DFB optical structures provide a fertile
classical simulator of non-Hermitian relativistic wave equations.
This work suggests that well-known phenomena occurring in DFB
structures, like spectral narrowing of resonances and
self-oscillation, are the measurable quantities associated to the
onset of spectral singularities and $\mathcal{PT}$ symmetry breaking
of Dirac Hamiltonians with certain complex couplings.\\
\\
Work supported by the italian MIUR (Grant No. PRIN-2008-YCAAK,
"Analogie ottico-quantistiche in strutture fotoniche a guida
d'onda").


\begin{thebibliography}{31}


\bibitem{Bender98}
C.M. Bender and S. Boettcher, Phys. Rev. Lett. {\bf 80}, 5243 (1998)

\bibitem{Bender02}
C. M. Bender, D.C. Brody, and H. F. Jones, Phys. Rev. Lett. {\bf
89}, 270401 (2002).

\bibitem{Mostafazadeh02}
A. Mostafazadeh, J. Math. Phys. {\bf 43}, 2814 (2002).

\bibitem{BenderReview}
C.M. Bender, Rep. Prog. Phys. {\bf 70}, 947 (2007).

\bibitem{MostafazadehReview}
A. Mostafazadeh, e-print arXiv:0810.5643.

\bibitem{RotterReview}
N. Moiseyev, Phys. Rep. {\bf 302}, 211 (1998); J.G. Muga, J.P.
Palao, B. Navarro, and I.L. Egusquiza, Phys. Rep. {\bf 395}, 357
(1998); I. Rotter, J. Phys. A {\bf 42}, 1 (2009).

\bibitem{Berry}
M.V. Berry, Czech. J. Phys. {\bf 54}, 1039 (2004).

\bibitem{O3}
 S. Klaiman, U. G\"{u}nther, and N.
Moiseyev, Phys. Rev. Lett. {\bf 101}, 080402 (2008).

\bibitem{O4}
A. Mostafazadeh, Phys. Rev. Lett. {\bf 102}, 220402 (2009).

\bibitem{D1}
C. Mudry, B.D. Simons, and A. Altland, Phys. Rev. Lett. {\bf 80},
4257 (1998); H. Egrifes and R. Sever, Phys. Lett. A {\bf 344}, 117
(2005); A. Sinha and P. Roy, Mod. Phys. Lett. A {\bf 20}, 2377
(2005); C.S Jia and A. de Souza Dutra, J. Phys. A {\bf 39}, 11877
(2006); F. Cannata and A. Ventura, Phys. Lett. A {\bf 372}, 941
(2008); O. Mustafa, S.H. Mazharimousavi, Int. J. Theor. Phys. {\bf
47}, 1112 (2008).

\bibitem{D2}
C.-S. Jia and A. de Souza Dutra, Ann. Phys. {\bf 323}, 566 (2008);
C-S. Jia, P.-Q. Wang, J.-Yi Liu, and S. He, Int. J. Theor. Phys.
{\bf 47}, 2513 (2008).

\bibitem{D3}
 V.G.C.S. dos Santos, A. de Souza Dutra, and
M.B. Hott, Phys. Lett. A {\bf 373}, 3401 (2009); F. Cannata and A.
Ventura, J. Phys. A {\bf 43}, 075305 (2010).

\bibitem{field}
C.M. Bender, K.A. Milton, and Van M. Savage, Phys. Rev. D {\bf 62},
085001 (2000); C.M. Bender, D.C. Brody, and H.F. Jones, Phys. Rev.
Lett. {\bf 93}, 251601  (2004); C.M. Bender, S.F. Brandt, J.-H.
Chen, and Q. Wang, Phys. Rev. D {\bf 71}, 065010 (2005); A.
Mostafazadeh, Int. J. Mod. Phys. A {\bf 21}, 2553 (2006).

\bibitem{O1}
A. Ruschhaupt, F. Delgado, and J. G. Muga, J. Phys. A {\bf 38}, L171
(2005).

\bibitem{O2}
R. El-Ganainy, K. G. Makris, D. N. Christodoulides, and Z. H.
Musslimani, Opt. Lett. {\bf 32}, 2632 (2007); K.G. Makris, R.
El-Ganainy, D.N. Christodoulides, and Z.H. Musslimani, Phys. Rev.
Lett. {\bf 100}, 103904 (2008); S. Longhi, Phys. Rev. Lett. {\bf
103}, 123601 (2009).

\bibitem{Salamo09}
A. Guo, G.J. Salamo, D. Duchesne, R. Morandotti, M. Volatier-Ravat,
V. Aimez, G. A. Siviloglou, and D. N. Christodoulides, Phys. Rev.
Lett. {\bf 103}, 093902 (2009).

\bibitem{Ruter10}
C.E. R\"{u}ter, K.G. Makris, R. El-Ganainy, D.N. Christodoulides, M.
Segev, and D. Kip, Nature Phys. {\bf 6}, 192 (2010).

\bibitem{Shank}
H. Kogelnik and C.V. Shank, J. Appl. Phys. {\bf 43}, 2327 (1972).

\bibitem{Poladian}
J. Carroll, J. Whiteaway, and D. Plumb, {\it Distributed feedback
semiconductor lasers} (The Institution of Electrical Engineers,
London, 1998).

\bibitem{note0}
A.B. Aceves and S. Wabnitz, Phys. Lett. A {\bf 141}, 37 (1989); D.N.
Christodoulides and R.I. Joseph, Phys. Rev. Lett. {\bf 62}, 1746
(1989).

\bibitem{Naimark}
M. A. Naimark, {\it Linear differential operators} (Ungar, New York,
1968); I.-P. P. Syroid, Ukrainian Math. J. {\bf 38}, 305 (1986);
G.S. Guseinov, Pramana {\bf 73}, 587 (2009).

\bibitem{note1}
Unitarity of the scattering matrix, implying $|t|^2+|r^{(l,r)}|^2=1$
and thus bounded values of transmission and reflection coefficients,
is satisfied for Hermitian Hamiltonians, however in non-Hermitian
systems unitarity may be broken [see, for instance: F. Cannata,
J.-P. Dedonder, and A. Ventura, Ann. Phys. {\bf 322}, 397 (2007)].

\end{thebibliography}

\end{document}